\documentclass[journal=jpclcd,manuscript=letter]{achemso}

\usepackage{microtype}
\usepackage{booktabs}
\usepackage{lmodern}
\usepackage{array}
\usepackage{listings}
\usepackage{mathpazo}

\usepackage[utf8]{inputenc}
\usepackage[ngerman,english]{babel}
\usepackage[T1]{fontenc}
\usepackage{graphicx}
\usepackage{amsfonts}
\usepackage{amsmath}
\usepackage{amssymb}
\usepackage{subcaption}
\usepackage{tabularx}
\usepackage{float}
\usepackage{icomma}
\usepackage{setspace}
\usepackage{placeins}
\usepackage{siunitx}
\usepackage{chemfig}
\usepackage{xcolor}
\usepackage{comment}

\title[]
{Polarization Resolved Optical Excitation of Charge-Transfer Excitons in PEN:PFP Co-Crystalline Films: Limits of Non-Periodic Modelling}

\author{Darius G\"under}
\affiliation{ Philipps-Universit\"{a}t Marburg, Molekulare Festk\"{o}rperphysik, 35032 Marburg, Germany}
\altaffiliation{Contributed equally to this work}

\author{Ana M. Valencia}
\affiliation{Carl von Ossietzky Universit\"at Oldenburg, Institute of Physics, Carl-von-Ossietzky-Stra{\ss}e 9, 26129 Oldenburg, Germany}
\altaffiliation{Contributed equally to this work}

\author{Michele Guerrini}
\affiliation{Carl von Ossietzky Universit\"at Oldenburg, Institute of Physics, Carl-von-Ossietzky-Stra{\ss}e 9, 26129 Oldenburg, Germany}

\author{Tobias Breuer}
\affiliation{ Philipps-Universit\"{a}t Marburg, Molekulare Festk\"{o}rperphysik, 35032 Marburg, Germany}

\author{Caterina Cocchi}
\affiliation{Carl von Ossietzky Universit\"at Oldenburg, Institute of Physics, Carl-von-Ossietzky-Stra{\ss}e 9, 26129 Oldenburg, Germany}
\alsoaffiliation{Humboldt-Universit\"{a}t zu Berlin, Physics Department and IRIS Adlershof, Zum Gro{\ss}en Windkanal 2, 12489 Berlin, Germany}
\email{caterina.cocchi@uni-oldenburg.de}

\author{Gregor Witte}
\affiliation{ Philipps-Universit\"{a}t Marburg, Molekulare Festk\"{o}rperphysik, 35032 Marburg, Germany}
\email{gregor.witte@physik.uni-marburg.de}

\begin{document}

\selectlanguage{english}

\newpage

\section*{Abstract}

Charge-transfer excitons (CTX) at organic donor/acceptor interfaces are considered important intermediates for charge separation in photovoltaic devices. Crystalline model systems provide microscopic insights into the nature of such states as they enable microscopic structure-property investigations. Here, we use angular-resolved UV/Vis absorption spectroscopy to characterize the CTX of crystalline pentacene:perfluoro-pentacene (PEN:PFP) films allowing to determine the polarization of this state. This analysis is complemented by first-principles many-body calculations, performed on the three-dimensional PEN:PFP co-crystal, which confirm that the lowest-energy excitation is a CTX. Analogous simulations performed on bimolecular clusters are unable to reproduce this state. We ascribe this failure to the lack of long-range interactions and wave-function periodicity in these cluster calculations, which appear to remain a valid tool for modelling properties of organic materials ruled by local intermolecular couplings.

\begin{figure}
	\centering
	\caption*{\textbf{TOC Graphic}}
	\includegraphics[width=5cm]{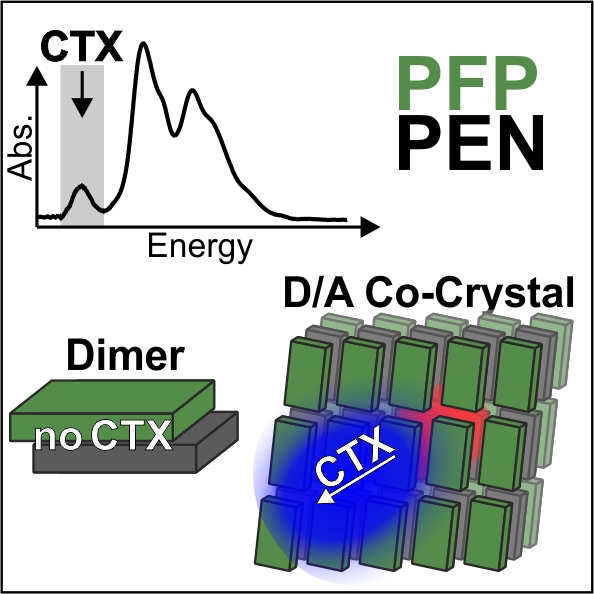}
	\label{img:TOC}
\end{figure}


\clearpage
\newpage

The large photo-absorption cross section of $\pi$-conjugated molecular materials favors their use for future thin film photovoltaic devices.\cite{jail+12nm,inga18am} 
However, light-matter coupling in these systems differs significantly from that of inorganic crystals, and the understanding of the underlying processes is by far not complete.\cite{razy+11se} 
In fact, in contrast to high performance inorganic solar cells, which are based on highly crystalline semiconductors, donor-acceptor blends are commonly used for the molecular counterparts. While this enhances the device efficiency by increasing the internal interface area,\cite{facc11cm,brab+11csr} their mostly amorphous structure hinders a microscopic analysis of the involved photophysical processes at the interfaces. Another reason is the complexity of the photo-physics of organic materials. Although pioneering theoretical schemes such as the H\"uckel theory or Kasha's model~\cite{kash+65} have significantly helped to rationalize the electronic structure and the optical transitions of the organic film constituents (see, \textit{e.g.}, Refs.~\citenum{anth06cr,span-silv14arpc} for review), these methods are \textit{per se} insufficient to address the full complexity of these materials. Since the turn of the century, \textit{ab initio} calculations have tremendously contributed to understand the electronic transport, and optical properties of organic semiconductors,~\cite{buss+02apl,cuda+12prb,rang+16prb,cocc+18pccp} and, in combination with model Hamiltonians, they have offered unprecedented insight into the excitations of these systems~\cite{delc+17prb,macd+20mh}, including electron-phonon coupling.~\cite{coro+02prb,hann+04prb}

Charge-transfer (CT) excitons at donor-acceptor interfaces are considered prime intermediates for electron-hole pair dissociation and free-carriers generation.~\cite{macd+20mh} To enhance the understanding of the microscopic nature of CT excitons in molecular composites, crystalline oligomeric model systems, especially organic co-crystals, are of particular interest, as they provide precise structural information on internal interfaces between the various components, which can be used to obtain correlations between structural and optoelectronic properties. While molecular cluster models allow to reproduce qualitatively some of the key features of these materials,~\cite{zimm+11jacs,arvi+20jpcb,mans+20jmcc,theu+21jpcc} a full-fletched computational analysis of the electronic and optical response of co-crystals with appropriate state-of-the-art methods still represents a numerical challenge. 

\begin{figure}
	\centering
	\includegraphics[width=0.5\textwidth]{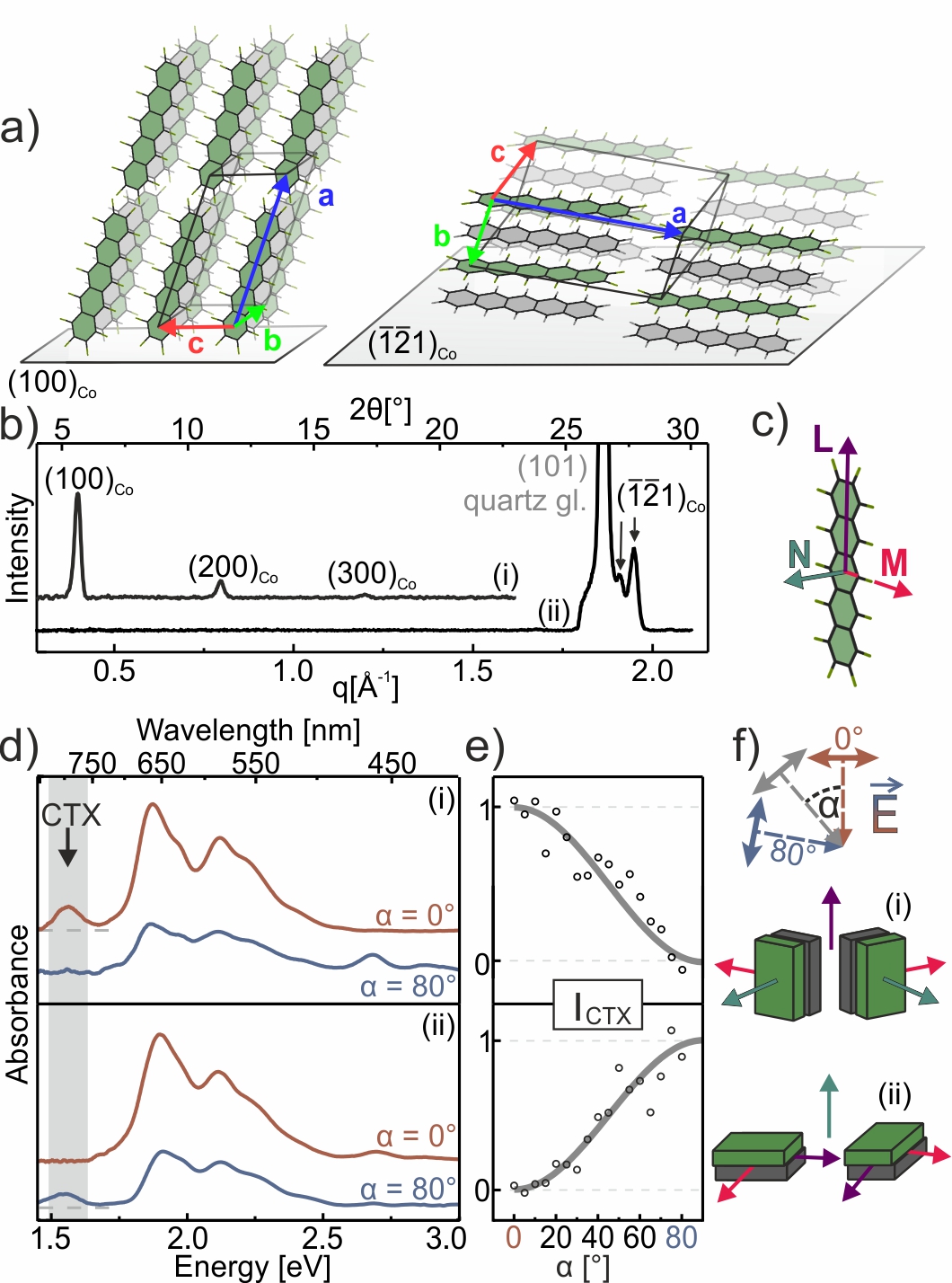}
	\caption{ (a) Structure of co-crystalline PEN:PFP films (PEN: grey, PFP: green) (b) Specular X-ray diffractograms of (i) a co-evaporated PEN:PFP film ($d_{nom}$=45nm, T=330K, glass substrate) with upright molecular orientation and of (ii) an alternating PEN:PFP monolayer stack of 75 double layers ($d_{nom}$=45nm, T=275K, graphene/quartz glass) (c) Definition of the molecular M-, N- and L-axis.  (d) Optical pathway corrected UV/vis transmission absorption spectra of both heterostructures, taken at different angles $\alpha$ of incidence of the linearly polarized light as depicted schematically in (f). (e) Normalized intensity of the CT-exciton absorption peak $I_{CTX}$ at 1.55eV in dependency on $\alpha$.}
	\label{Figure:fig1}
\end{figure}

Among polycyclic hydrocarbons, acenes are known to form
well-ordered crystalline films.\cite{anth06cr} Their typical p-type behavior can be chemically converted to n-type by fluorination,\cite{tang-bao11cm} hence providing donor-acceptor pairs without changing molecular shape or size. In particular, for pentacene (PEN) and perfluoropentacene (PFP),\cite{sako+04jacs} the optoelectronic properties of unitary solid films have been extensively studied.\cite{hind+07jpc, breu-witt11prb,smit-mich13arpc,kola+04acsn,cocc+18pccp} In the case of PEN:PFP blends and layered crystalline 
heterostructures, a new low energetic state was found in the absorption spectra and assigned to a CT exciton but without identification of its molecular character.\cite{broc+11prb,ange+12jcp,breu-witt13jcp} In subsequent works, also the lifetime and decay dynamics of this CT exciton as well as its feeding from singlet excitons of acceptor molecules in heterostructures were examined.\cite{drinn+17ami,broc+17pssrrl,kim+19jcp,hans+21ami} Unfortunately, the low solubility and diffusivity of PEN and PFP have hindered the growth of films with sufficiently large single-crystalline domains, and, hence,
polarization-resolved optical measurements on individual domains, which are possible for unitary acene films.\cite{breu-witt11prb,kola+04acsn,cocc+18pccp,helt+11apl,gund+20ami} However, the crystal structure of equimolar polycrystalline PEN:PFP films, prepared either by co-evaporation onto SiO$_2$ or by alternate monolayer deposition onto graphite, has recently been resolved. Interestingly, both film preparations yield essentially the same crystal structure, but reveal different molecular orientations, see Figure~\ref{Figure:fig1}a: standing molecules on SiO$_2$ and almost flat-lying molecules on graphite,\cite{davi+20cm} which parallels the previous template-controlled growth of acene heterolayers.~\cite{breu-witt15acsami}

In this study, we used such differently oriented co-crystalline PEN: PFP films to characterize the polarization of the optical excitations and found that the transition dipole moment (TDM) of the low energy excitons is oriented along the stacking direction. In parallel, we perform a first-principles many-body analysis of the optical response of these materials. 
Our results demonstrate that neither the application of Kasha's model,\cite{kash+65}
which was successfully used to describe the polarization of the individual Davydov components of excitons in unitary acene films,~\cite{gund+20ami} nor modeling the PEN:PFP films as isolated bimolecular clusters lead to the correct polarization of the low energy excitation. 
Only the treatment of the system as a periodic co-crystal restores the experimentally observed picture and thereby confirms the CT character of the low energy exciton, thus emphasizing the crucial role of long-range electronic correlations in reproducing these phenomena.~\cite{guer+21jpcc}

To enable polarization resolved UV/Vis spectroscopy on co-crystalline solids, we followed the approach of D’Avino \textit{et al.}~\cite{davi+20cm} and prepared different PEN:PFP heterostructures of equimolar stoichiometry by (i) co-deposition onto a glass slide and (ii) alternating deposition of PEN and PFP monolayers onto a graphene-coated quartz glass substrate. 
As depicted in Figure~\ref{Figure:fig1}b, the corresponding X-ray diffractograms reveal in the first case only (n00) reflections of the co-crystalline structure corresponding to a lattice spacing of 14.6~\AA{} and thus confirm an uniform upright molecular orientation. In contrast, the specular X-ray diffractogram of the second heterostructure reveals no (n00) reflection, indicating the absence of any crystalline regions with upright molecular orientation. Instead, the specular diffractogram is dominated by an intense (101) reflection of the quartz glass substrate at q=1.86~\AA{}$^{-1}$ (2$\theta=26.5^{\circ}$), but also shows two additional weak reflections at q=1.914~\AA{}$^{-1}$ (2$\theta=27.1^{\circ}$) and q=1.952~\AA{}$^{-1}$ (27.7$^{\circ}$). 
Such reflections are different to those of unitary phases~\cite{salz+12acsn,breu-witt15acsami,gund+20ami} and belong to the co-crystalline phase, thus proving a complete intermixture of PEN and PFP on a molecular level.
These signals are assigned to $\mathrm{(\overline{1}\overline{2}1)}$ reflections of two slightly differing co-crystalline phases and correspond to an interlayer spacing of 3.2~\AA{}, which is in perfect agreement with the previous study by D’Avino \textit{et al.} and thereby confirms an exclusive recumbent molecular orientation.  

With this precise knowledge about the co-crystalline arrangement in PEN:PFP films, the polarization resolved macroscopic optical characterizations performed in transmission geometry can also be related to the molecular M-, N- and L-axes (cf. Figure~\ref{Figure:fig1}c). As depicted in Figure~\ref{Figure:fig1}d, film (i) exhibits various bands above 1.8~eV, which are assigned to interband transitions
and vibronic progressions (SI, Section S3). Additionally, a weak band is present at 1.55~eV, which has also been resolved in previous works, where it was attributed to a CT exciton.
\cite{broc+11prb,ange+12jcp, breu-witt13jcp} Systematic variation of the preparation conditions shows that homogeneous mixing of such uprightly oriented molecules requires substrate temperatures of 300K or more upon growth and is substantially reduced for lower temperatures, indicated by a lower intensity of the CT-exciton signal while no defect related signatures are observed in the absorption spectra (for further discussion on the role of defects see section S5 of the Supporting Information). The weak but distinct CT-exciton-related absorption in the co-crystalline blend prepared at 330K shows that this excitation has a non-vanishing oscillator strength. To obtain additional information about the orientation of its TDM, absorption spectra are recorded at different angles of incidence, allowing to vary the angle between the incident electric field (E) and the sample surface (Figure~\ref{Figure:fig1}d-f). For the quantitative analysis linearly p-polarized light is used and the enhanced pathway due to the sample inclination is considered. This analysis scheme is validated for unitary acene films with uniform molecular orientation (both standing and lying) whose TDM orientations are well-known (see SI, Figures~S3-S4). 

Figures~\ref{Figure:fig1}d,e show a clear variation of the peak intensity of the CT exciton, which is maximal at normal incidence ($\alpha= 0^{\circ}$) while it vanishes for grazing illumination ($\alpha > 80^{\circ}$) of film (i) and can be described by a $cos^2(\alpha)$ function. Based on the general expression of optical dipole excitations, I$_{Abs.}$  $\propto$ | $\vec{E} \cdot \vec{(TDM)}$|$^2$, we can conclude that the TDM of the CT exciton is oriented parallel to the sample surface. However, due to the presence of azimuthally isotropically distributed  crystalline domains in such films, it cannot be distinguished whether the TDM is oriented along the molecular N- or M-axis since they are both oriented within the surface plane [see scheme (i) of Figure~\ref{Figure:fig1}f]. Additional information is provided by the optical signature of the heterostructure (ii) with recumbent molecular orientation. In that case, a substrate temperature of 275K is sufficient to enable spontaneous intermixing of recumbently oriented PEN and PFP molecules, as previously found also for a mixed acene monolayer on $\mathrm{MoS_2}$,\cite{kach+21cs} while partial dewetting typically occurring at elevated growth temperatures is largely suppressed. The corresponding optical absorption spectrum at normal incidence (Figure~\ref{Figure:fig1}d(ii), $\alpha= 0^{\circ}$) reveals no signatures of the CT exciton, which becomes visible at increasing incidence angle and the resulting angular-dependent excitation can be well described by a $\mathrm{sin^2}(\alpha )$ function (see Figure~\ref{Figure:fig1}e). At grazing incidence, the E-field of the illuminating light is oriented essentially parallel to the molecular N-axes. We note that only by combining the results of the two films with different molecular orientations we can conclude that the TDM of the CT exciton in co-crystalline PEN:PFP films must be oriented close to the stacking direction.

\begin{figure}[h!]
	\centering
	\includegraphics[width=\textwidth]{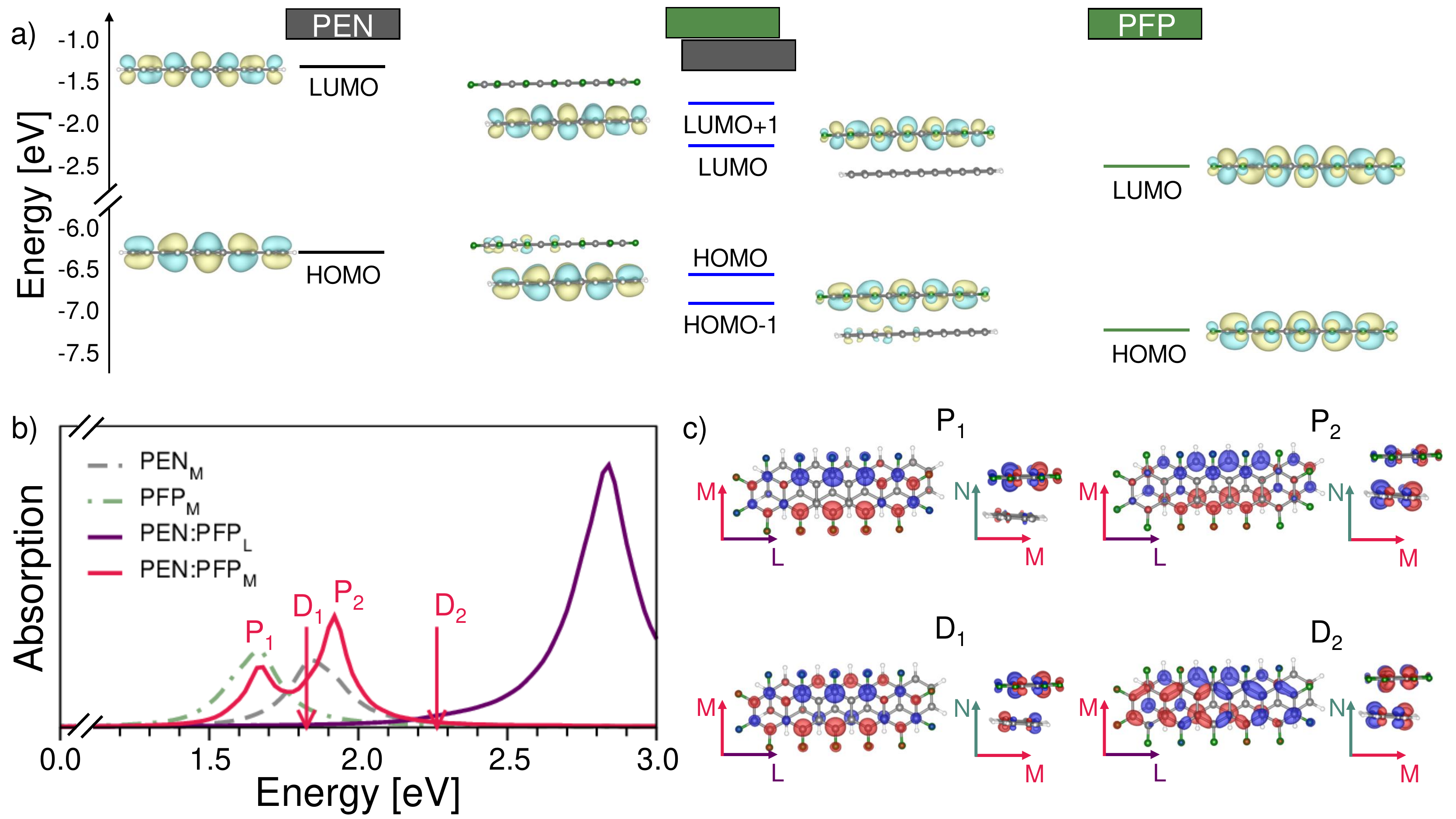}
	\caption{(a) Energy levels and molecular orbitals of PEN, PEN:PFP, and PFP obtained from $G_0W_0$ on top of CAM-B3LYP. (b) BSE spectra of PEN (dashed line), PFP (dashed-dotted line), as well as the PEN:PFP cluster for both polarizations along the long (L) and short (M) molecular axes (solid lines). A Lorentzian broadening of 150 meV is applied to all spectra. (c) Transition densities of the bright (P$_1$ and P$_2$) and dark excitations (D$_1$ and D$_2$) in the PEN:PFP cluster. }
		\label{fig:cluster}
\end{figure}

To rationalize the experimental findings and to identify the character of the low energy excitation, we perform first-principles 
calculations, based on density-functional theory (DFT) and many-body perturbation theory, initially considering a bimolecular cluster formed by a PEN and a PFP molecule with the respective basal planes facing each other, as in the unit cell of their co-crystal.~\cite{davi+20cm}
The molecular orbitals (MOs) of this system are alternately distributed on either molecule, with the HOMO and the LUMO+1 localized on PEN and the HOMO-1 and the LUMO on PFP (see Figure~\ref{fig:cluster}a).
These four orbitals correspond to the frontier orbitals of the isolated molecules (Figure~\ref{fig:cluster}a and Figure~S9 in the SI).
Hence, unsurprisingly, the optical absorption spectrum obtained for the 
two stacked molecules from $GW$/Bethe-Salpeter equation (BSE) calculations~\cite{brun+16cpc} on top of hybrid DFT,~\cite{yana+04cpl} resembles, at least at the onset, those 
of the isolated gas-phase molecules computed at the same level of theory (see Figure~\ref{fig:cluster}b).
Two peaks, labeled P$_1$ and P$_2$, appear at lowest energies in the spectrum of the cluster and correspond to the transitions from both the HOMO-1 and the HOMO to the LUMO, and from the HOMO to the LUMO+1, respectively (see Table~S1 in the SI). 
Both are intramolecular excitations, with P$_1$ mainly localized on PFP while P$_2$ on PEN (see Figure~\ref{fig:cluster}c).
The similarity between the absorption maxima in the spectrum of the cluster and in those of the isolated moieties is not limited to their energy, composition, and oscillator strength.
In all cases, the polarization of the excitations is along the short axis of the molecules (M), as indicated by the transition density (TD) plots (Figure~\ref{fig:cluster}c), providing information about the polarization direction and the spatial distribution of the excited states.
It is worth noting that two dark excitations, labeled D$_1$ and D$_2$, are found above P$_1$ and P$_2$, respectively.
None of them corresponds to a CT-exciton.
D$_1$ has a very similar composition as P$_1$ however, in this case, there is a destructive interference which can be seen from the opposite phases of the density domains in the two molecules (Figure~\ref{fig:cluster}c).
On the other hand, for D$_2$, the orientations of the dipole domains of the TD are on different directions in the two molecules (along L for PFP and along M for PEN).

At higher energies in the spectrum of the bimolecular cluster, between 2.5 and 3.0~eV (see Figure~\ref{fig:cluster}b), an intense peak is visible. 
This maximum has again a clear origin from the excitations of the individual molecules, which are polarized along their long molecular axis (L) in that energy region (see Table~S2 for PFP).
However, in the 
window displayed in Figure~\ref{fig:cluster}b, no signatures compatible with the charge-transfer excitons observed in the experimental spectra appear. Although the energy of P$_1$ is close to the energy of the CT-exciton, the TDM of P$_1$ is oriented along the M-axis and thus, P$_1$ correlates rather to the second lowest band observed experimentally. 
To exclude that this discrepancy is related to the specific arrangements of the molecules in the cluster, we additionally considered a bimolecular cluster in which the planes of PEN and PFP are aligned laterally almost along their short axes (see SI, Figure~S9c).
The results obtained for this system are consistent with those reported in Figure~\ref{fig:cluster} and confirm that, regardless of the  molecular packing motif, molecular cluster models are \textit{per se} inadequate to reproduce the spectral fingerprints of the co-crystal. 
Based on the results from previous work,~\cite{vale+20pccp,theu+21jpcc} where also larger stacks of donor/acceptor molecules were examined at the same level of theory adopted here, we can exclude that increasing the number of molecules in the cluster will change the picture provided by this model. 
The reason for this deficiency is related to the absence of long-range intermolecular interactions~\cite{guer+21jpcc} promoting excitations polarized along the stacking direction.
To correctly reproduce the optical features of the co-crystal, the periodicity of the system must be taken explicitly into account.

\begin{figure}[h!]
	\centering
\includegraphics[width=1.0\textwidth]{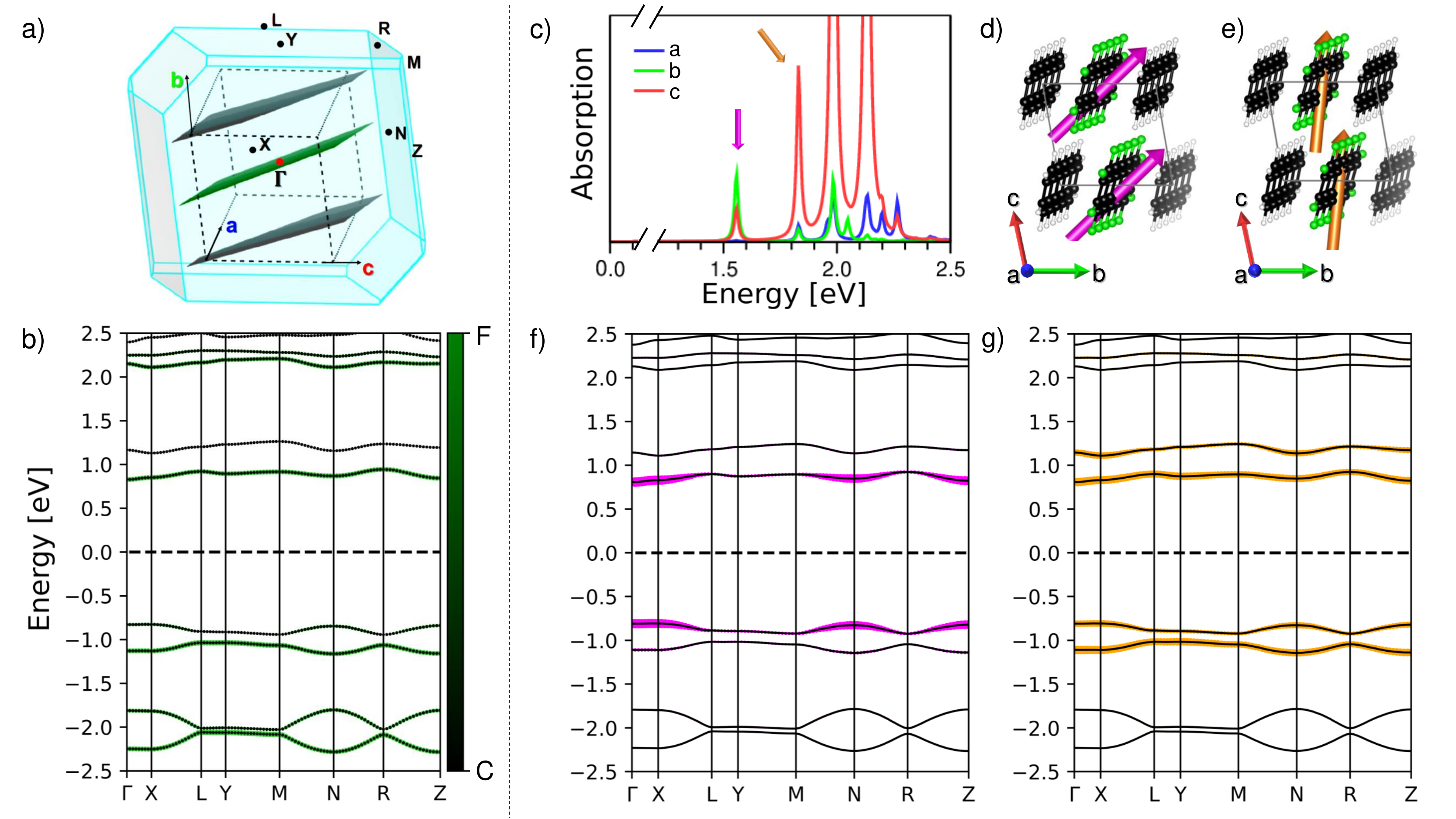}
	\caption{a) Brillouin zone (BZ) of the PEN:PFP co-crystal with superimposed real space unit cell. b) Single-particle band structure of the PEN:PFP co-crystal with orbital contributions from C-and F-atoms. 
	c) BSE absorption spectrum resolved in three components along the crystal axes. A Lorentzian broadening of 10 meV is included. The colored arrows mark the first and second bright excitations analyzed in panels d) and e) in terms of TDM vectors, and in f) and g) in terms of exciton weights (see SI, Section S6). 
	} 
	\label{fig:optic}
\end{figure}

For this purpose calculations were performed using the previously resolved co-crystalline structure~\cite{davi+20cm} (cf. Figure \ref{Figure:fig1}a, \ref{fig:optic}a).
The electronic structure of the PEN:PFP co-crystal  reflects its anisotropic geometry.
The valence and conduction bands around the gap are almost flat and are localized either on PEN or PFP~\cite{davi+20cm} (see Figure ~\ref{fig:optic}b). To visualize the band character we use orbital projections on C and F atoms, as in this energy region the PFP wave-functions are always spread also on the F-atoms (see Figure~S8).

The optical absorption spectrum computed for the PEN:PFP co-crystal is displayed in Figure~\ref{fig:optic}c with the tensor components projected along the crystalline axes.  
Due to the complex arrangement of the molecules in the unit cell, we visualize the TDM vectors associated to the first two bright excitons (Figure~\ref{fig:optic}d,e).
For a quantitative analysis of these excitons, we also display corresponding fat-band plots (see Figure~\ref{fig:optic}f,g and SI for further details).
The lowest-energy peak, marked by a magenta arrow in Figure~\ref{fig:optic}c, corresponds to a relatively weak excitation polarized mainly along $b$ and partly along $c$ (Figure~\ref{fig:optic}d), stemming from a few transitions around the high-symmetry points X, N, and Z between the highest occupied band and the lowest unoccupied one localized on the PFP and PEN molecule, respectively (Figure~\ref{fig:optic}f).
As such, this is a CT-exciton.
The second peak (golden arrow in Figure~\ref{fig:optic}c) corresponds to the third excitation (the second has no oscillator strength) and it is more intense than the first one. 
It is mainly polarized along the $c$ axis with minor contributions from components along $a$ and $b$, and, as such, corresponds to an intramolecular transition approximately polarized along the short molecular axis M (Figure~\ref{fig:optic}e).
It originates from transitions between the second highest-occupied band and the lowest-unoccupied one (Figure~\ref{fig:optic}g), both dominated by PFP states (Figure~\ref{fig:optic}b), with additional contributions coming also from the highest valence band and the second lowest conduction band.
The weights of the single-particle transitions of this excitation are much more delocalized in \textbf{k}-space compared to their counterparts in the first excitation (compare Figure~\ref{fig:optic}f and g), pointing to an enhanced real-space localization~\cite{cocc+18pccp,fu+17pccp} which confirms the intramolecular character of this excitation within the PFP molecules. 
At higher energies in the spectrum of Figure~\ref{fig:optic}c, two intense peaks appear: they are predominantly polarized along the $c$ axis and bear similar character as the second peak discussed above (for further details, see SI, Figure~S10).

The analysis of the optical excitations computed for the PEN:PFP co-crystal reported above demands an additional comment regarding the role of excitonic effects, which are embedded in the adopted first-principles formalism based on the solution of the BSE.~\cite{vorw+19es}
The comparison between the spectrum shown in Figure~\ref{fig:optic}c and its counterpart in which excitonic effects are neglected (Figure~S11 in the SI) reveals not only that the first maximum has CT character but that all the 
peaks up to 2.2~eV result from the inclusion of 
electron-hole interactions.
The different compositions in terms of single-particle transitions and the different spatial distribution of these 
states is typical of organic (co-)crystals, due to the complex interplay between intra- and intermolecular coupling.\cite{cocc+18pccp,guer+21jpcc} 
We emphasize that this insight is available only through 
the methodology adopted in this work.

The \textit{ab initio} results obtained
for the co-crystal considered in its full periodic arrangement are evidently very different, even qualitatively, from those given by an isolated bimolecular cluster \textit{in vacuo}. 
Only in the former case, good agreement with experimental data is achieved.
We can rationalize this finding as follows:
Calculations on isolated clusters are necessarily ruled by the physics of the individual molecular components.
As such, they are suited to describe level alignments~\cite{salz+12prl,theu+21jpcc,zeis+21jpcl} and other properties related to the local interactions between the moieties,~\cite{vale-cocc19jpcc,vale+20pccp} including the effect of an electrostatic environment, which induce, \textit{inter alia}, a renormalization of the ionization energy~\cite{orts+20afm,krum+21jcp} but does not alter the quantum-mechanical interactions among the molecules.
Indeed, we found that the lowest-energy excitations of the PEN:PFP cluster 
have approximately the same energy, oscillator strength, composition, and polarization along the molecular M-axis as their counterparts in gas-phase PEN and PFP. 
This expected finding is intrinsic to the non-periodic character of the adopted model independently of the chosen molecular arrangement (see SI, Figure S9).
Based on previous work on donor-acceptor co-crystals,~\cite{vale+20pccp} we anticipate that this finding is also independent of the number of molecules included in the cluster: as long as the system is treated as non-periodic, the electronic and optical response of the co-crystal cannot be correctly reproduced.

Simulating the co-crystal as a solid enables the inclusion of the physical ingredients that rule the problem: the periodicity of the wave-functions and the long range Coulomb interactions through the crystal lattice acting on them in the ordered and periodic molecular array.~\cite{guer+21jpcc}
Notice that, although less intense than the higher-energy maxima, the lowest-energy excitation in the co-crystal is optically active.
The presence of CT-excitons with non-zero oscillator strengths is a characteristic of extended materials, where wave-function delocalization ensures long-range spatial overlap.
We emphasize that these effects are intrinsic to the periodic treatment of the electronic structure of the material and are captured already in the mean-field framework of DFT. 
The explicit inclusion of electron-hole correlations in the calculation of optical spectra provide us with accurate results and with quantitative information about the composition of the excitations, their polarization, and also of their binding energies.
In the PEN:PFP co-crystal, the binding energy of the CT-exciton is approximately 0.8~eV, \textit{i.e.,} larger than in the unitary pentacene crystal.~\cite{cocc+18pccp}
The CT-character of this exciton is enhanced by the electron-hole interactions (see SI, Figure~S11).
In the whole spectrum, excitonic effects act through a redistribution of the oscillator strength to lower energies. 

In summary, by means of angular-resolved optical absorption spectroscopy, we could identify and determine the polarization of a CT-exciton at the absorption onset of well-defined co-crystalline PEN:PFP films.
Using first-principles many-body calculations, we demonstrated that this key characteristic of the material can be correctly reproduced and rationalized only by simulating the co-crystal as a three-dimensional solid.
Cluster calculations are unable to accomplish this task, due to the missing consideration of the periodic potential and of long-range interactions between the periodically arranged molecules. 
We are confident that our results will stimulate further analysis to disclose the impact of long-range interactions also in other relevant properties (\textit{e.g.,} charge transport) in organic (co-)crystals.


\section*{Acknowledgements}
D.G., T.B., and G.W. acknowledge financial support provided by the German Science Foundation (Deutsche Forschungsgemeinschaft, DFG) project-ID 223848855-SFB 1083 “Structure and Dynamics of Internal Interfaces” within the project A2. D.G. gratefully acknowledges financial support by the Cusanuswerk. A.M.V., M.G., and C.C. acknowledge financial support from the German Research Foundation, project number 182087777 (CRC 951), from the German Federal Ministry of Education and Research (Professorinnenprogramm III), and from the State of Lower Saxony (Professorinnen für Niedersachsen). The computational resources were provided by North-German Supercomputing Alliance (HLRN) through the project bep00076. A.M.V. thanks Christian Vorwerk and Ignacio Gonz\'alez for helpful discussions.

\begin{suppinfo}
Details on the experimental setup and computational methods as well as additional UV/Vis absorption measurements and calculations are provided in the Supporting Information. 
\end{suppinfo}



\providecommand{\latin}[1]{#1}
\makeatletter
\providecommand{\doi}
  {\begingroup\let\do\@makeother\dospecials
  \catcode`\{=1 \catcode`\}=2 \doi@aux}
\providecommand{\doi@aux}[1]{\endgroup\texttt{#1}}
\makeatother
\providecommand*\mcitethebibliography{\thebibliography}
\csname @ifundefined\endcsname{endmcitethebibliography}
  {\let\endmcitethebibliography\endthebibliography}{}

\end{document}